\documentstyle[12pt]{article}
\newcommand{\be}{\begin{eqnarray}}

\newcommand{\ee}{\end{eqnarray}}

\title{
        \begin{flushright}
        {\normalsize
        DOE/ER/40561--297--INT96--21--002\\
        TPI-MINN-97/17\\
        NUC-MINN-97/4--T\\
        NBI--97--08\\
        November 1997 \\}
        \end{flushright}
\bf     Boost covariant gluon distributions in large nuclei       
       }
\author{Larry McLerran\\ 
       {\small\it Theoretical Physics Institute, University of Minnesota, 
        Minneapolis, MN 55455} \\
        Raju Venugopalan \\
        {\small\it Niels Bohr Institute,
        Blegdamsvej 17,
        Copenhagen, Denmark, DK--2100 } \\          
       }

\date{}

\begin{document}
\setcounter{page}{0}
\maketitle
\thispagestyle{empty}
\begin{center}
{\bf Abstract}\\
\end{center}

\noindent 

It has been shown recently that there exist analytical solutions of the
Yang--Mills equations for non--Abelian Weizs\"acker--Williams fields 
which describe the distribution of gluons in large nuclei
at small x. These solutions however depend on the color charge distribution
at large rapidities. We here construct a model of the color charge 
distribution of partons in the fragmentation
region and use it to compute the boost covariant momentum distributions of 
wee gluons. The phenomenological applications of our results are 
discussed.

\vfill \eject

\section{Introduction}
\vspace*{0.3cm}

Recently, we constructed a QCD based model for the distribution of gluons
in large nuclei at small x~\cite{RajLar}. These fields are the non--Abelian
analog of the well known Weizs\"acker--Williams fields in QED. A path integral
was written down for the wee partons in the presence of valence quark 
sources. These sources are static charges on the light cone. In 
subsequent papers the saddle point solution of the path integral and the 
influence of quantum fluctuations about the saddle point solution were 
investigated~\cite{RajLar2}. It was shown later by one of us in Ref.~
\cite{AlexLar} (with J. Jalilian--Marian, A. Kovner and H. Weigert) that 
the validity of the original model could be extended to finite nuclei and even
protons at very small values of x. Remarkably, it was also shown by 
J. Jalilian--Marian et al.\cite{AlexLar}  and by Kovchegov \cite{Kovchegov}
that the classical Yang--Mills equations for the 
non--Abelian Weizs\"acker--Williams fields can be solved and an analytical 
solution obtained for the nuclear gluon distribution at small values of x.

However, the gluon distributions at a particular value of x depend on the 
color charge squared per unit area, $\chi(y, Q^2)$, which is the integral
of the color charge squared at all rapidities greater than the rapidity
of interest. Formally, we define $\chi$ to be
\be
\chi (\eta, Q^2) = \int_\eta^{\infty} d\eta^\prime\, \mu^2 (\eta^\prime, Q^2) \, ,
\ee
where $\mu^2 (\eta^\prime, Q^2)$ is the color charge squared per unit area
per unit rapidity at the rapidity $\eta^\prime$ and $Q^2 >> \Lambda_{QCD}^2$ 
is the momentum transfer squared of an external probe. Note also that
the space--time rapidity is defined by the relation 
\be
\eta=\eta_{R} + \log(x_{R}^-/x^-)\, .
\label{raps}
\ee
Here $x_{R}^{-}\approx 2R/{\sqrt{2}\gamma}$ is the Lorentz contracted 
thickness 
of the nucleus in the infinite momentum frame ($\gamma >> 1$) and 
$x^-=(t-z)/\sqrt{2}$ is the longitudinal light cone co--ordinate. At central
rapidities, the space--time rapidity is approximately equal to the momentum
space rapidity.

In order to explicitly compute the parton distributions
in nuclei at small x, we need to model the color charge distribution of 
partons in the fragmentation region. In this paper we will construct a 
simple model which captures the essential physics of the fragmentation 
region in nuclei. The expression for $\chi$ that results will lead to a boost 
covariant expression for the gluon distribution in large nuclei at small x. 
A conceptual issue that is resolved as a consequence is the description of 
the non--Abelian Weizs\"acker--Williams fields in the nuclear rest frame.
This enables us in principle to compare our results to calculations in
the nuclear rest frame~\cite{BrodHeb}.

In large part, this work is an extension of the nice analysis done by 
Kovchegov \cite{Kovchegov} who, in a simplified model of a nucleus, formulated
in the nuclear rest frame, argued
that the nucleus acts as a source of color charge.  Here we incorporate a
model of the longitudinal dynamics and construct a solution in the 
fragmentation region of the nucleus.  We argue that in most of the 
fragmentation region for a sufficiently large nucleus, our
weak coupling methods are valid.

We should warn the reader, that our results for realistically sized nuclei
are semi-quantitative and qualitative at best.  For realistically sized nuclei,
the typical energy scale in the problem  is of the order of $300$--$400$ MeV at
RHIC energies and $750$--$900$ MeV at LHC energies at central rapidities~\cite
{gyulassy},
which is marginally large enough (for our weak coupling calculations to
be valid) only in the case of the latter. 
In order to get a scale well in the weak coupling regime would require a 
nuclear baryon 
number a couple of orders of magnitude larger than that of the largest 
nucleus or equivalently, much higher energies.
Nevertheless, we believe the insight we gather from this picture may
be useful for obtaining a reasonable, QCD motivated conceptual framework in 
which to think about high energy nuclear collisions.

Our results also enable us to study the sytematics of nuclear gluon shadowing
and other phenomenological observables at low x. These will be especially 
relevant for the deep inelastic scattering experiments off nuclei which have
been proposed at HERA~\cite{HERA}. A caveat is that our result 
for the gluon distribution is classical. 
However, another important result of Jalilian--Marian et al.~\cite{AlexLar}
is that the effective action for wee partons demonstrates a self--similar
structure as we go to lower rapidities. The effective 
color charge $\chi$ obeys a renormalization group equation analogous to the
DGLAP or BFKL equations~\cite{BFDGL} (we note that in a recent paper, it has been shown explicitly that the BFKL kernel is recovered in this approach~\cite{Kovner}). Therefore, at each rapidity slice,
the gluon distribution is given by the classical gluon distribution with the
quantum effects absorbed in the effective scale $\chi$ at that rapidity slice.
The residual quantum corrections are of order $\alpha_S\Delta \eta$ which is
much smaller than unity for the rapidity slice of width $\Delta \eta$. 
These can in 
principle be computed. In other words, the gluon distributions computed using 
our model for $\chi$, provide the initial conditions for the renormalization 
group equations which can be solved self--consistently for $\chi$.

The rest of the paper is organized as follows. In section 2, we briefly 
discuss and summarize 
the results of Ref.~\cite{AlexLar} for the classical gluon distribution. 
In section 3, we construct our model for $\chi$, the color charge squared 
per unit area. Our model depends on one free parameter, $\kappa$, which is the
average color charge squared per {\it nucleon} integrated over one unit of 
rapidity. Section 4 contains a discussion of how $\kappa$ may be determined
and the implications of our model for shadowing in very large nuclei. 
The final section contains a brief summary and points out directions for 
future work.

\section{An analytical solution for correlation functions of non--Abelian 
Weizs\"acker--Williams fields}
\vspace*{0.3cm}

In Ref.~\cite{RajLar,AlexLar}, a path integral was written down for wee parton 
modes in a nucleus with $x<<A^{-1/3}$ and transverse momentum 
$\Lambda_{QCD}<< k_t<< \sqrt{\chi}$. Here $\chi=\int_\eta^{\infty} 
d\eta^\prime 
\mu^2(\eta^\prime,Q^2)$, where $\mu^2$ is the 
average color charge squared per unit area at a particular rapidity 
$\eta^\prime$
and a transverse momentum resolution $Q^2$. The latter is normally associated
with the momentum resolution of an external probe. The
measure for these modes (in light cone gauge $A^+=0$) is
\be
        Z & = & \int~ [dA_t \, dA^-] [d\rho]
\exp\Bigg( iS_{QCD} 
- ig\int d^4x A^-(x)\rho (x_t,\eta) \nonumber \\
 &-& \int_0^\infty d\eta 
\int d^2x_t {1 \over {2\mu^2 
(\eta,Q^2)}} \rho^2 (\eta,x_t) \Bigg) \, .
\label{funp}
\ee
Above, $y$ has the same definition as in Eq.~\ref{raps}. In the trace that
defines the path integral, the sum over the valence quark color charges has
been replaced by a Gaussian measure with the weight $\mu^2$. The action 
$S_{QCD}$ above refers only to the pure gauge degrees of freedom. The above
action does not include sea quarks (which are $\alpha_S$ suppressed) but
is easily extended to do so. Since $\mu^2$
is the only scale in the problem, the coupling constant will run as a 
function of this scale. For $\mu^2 >> \Lambda_{QCD}^2$, our effective theory
will be in the regime of weak coupling.

This procedure is justified when the following two assumptions apply~\footnote{For a recent discussion of a gauge invariant form for the above action, see
Ref.~\cite{Kovner}. This discussion while of general interest, is not applicable here since we restrict ourselves to the semi--classical result.}. The 
first is that the valence quarks are static 
charges on the light cone with $J^{+,a}=\rho^a (x_t)\delta(x^-)$ as the only 
large component of the valence quark current. All other components of the
current are suppressed by $1/P^+$. The second assumption is that the wee 
partons couple to a large color charge which is classical. What this implies
is that the charge squared per unit area, $<Q^2>$, that the wee partons
couple to is given by the relation $<Q^2>=N_q <Q^2>_{quark}$, where $N_q$ 
is the number of quarks per unit area ($\propto A^{1/3}$) 
and $<Q^2>_{quark}$ is the charge squared
per unit area due to a single quark. It can be shown that the central limit
theorem can be applied to this problem and that the corrections to the 
classical result (which may in principle be computed) are of the order
$1/\sqrt{N_q}\sim A^{-1/6}$ for the root mean square classical charge~\cite{
RajNex}. For an alternative derivation of the Gaussian measure in a specific
model, see Ref.~\cite{Kovchegov}.

To compute the classical nuclear gluon distribution function
\be
{dN\over {d^3 k}} = {2\over {(2\pi)^3}} 2 |k^+| \int d^3 x d^3 x^\prime
e^{ik\cdot (x-x^\prime)} <A_i^a(x^-,x_t)A_j^b({x^\prime}^-,x_t^
\prime)>_{\rho} \,\, ,
\label{Green}
\ee
one needs to compute the saddle point solution of Eq.~\ref{funp},
or equivalently, solve the classical Yang--Mills equations for the source
$J^+$ defined above and finally average over the product of the classical 
fields at two space--time points with the Gaussian measure in Eq.~\ref{funp}. 
Note that the brackets $<\cdots>_{\rho}$ above correspond to the Gaussian 
averaging over the static, light cone sources. 
One finds a particular 
solution where $A^{\pm}=0$ and
$A^i (i=1,2)$ is a pure gauge field which satisfies the equation
\be
D_i {d A^i\over d\eta} = g \rho (\eta,x_\perp) \, ,
\ee  
where $D_i$ is the covariant derivative~\cite{RajLar}. 

In Ref.~\cite{AlexLar},
the above equations were solved to obtain $A^i$ as
a function of the color charge density $\rho$ and the distribution function
computed.
Here we will merely state the result of Ref.~\cite{AlexLar} and refer the
reader to that paper for the technical details. The distribution function
for the non--Abelian Weizs\"acker--Williams fields in coordinate space is
\be
{dN\over {d^2 x_t d\eta}}={4 \over g^2}
{(N_c^2-1)\over {N_c x_t^2}}\left(1-\left(x_t^2
\Lambda_{QCD}^2\right)^{{g^4 N_c\over {16\pi}}\chi(\eta,Q^2)x_t^2}\right) \, .
\label{spaceglue}
\ee

The momentum space distributions are easily obtained by Fourier transforming
Eq.~\ref{spaceglue}. The result is
\be
{1\over {\pi R^2}}\,{dN\over {dx_A d^2 k_t}}&=& {4\over{\pi^2}}\,{(N_c^2-1)
\over N_c}\,(r_0 m_N)^2 {1\over {4\pi\alpha_S}} x_A 
 \int_0^{\log(1/x_A)} dy^\prime \exp(2y^\prime)\nonumber \\
&\times&
\int_0^{1/\Lambda_{QCD}} {du_\perp\over u_\perp} J_0 (k_t u_\perp)\,
\left[1-(u_\perp \Lambda_{QCD})^\beta\right] \, .
\label{fourier}
\ee
Above, $r_0=1.12$ fm, $m_N$ is the nucleon mass and $\beta = 2\pi \alpha_S^2 
N_c \chi(y^\prime,Q^2)u_{\perp}^2$. For $\alpha_S\sqrt{\chi}<<k_t$, the distribution has
the familiar Weizs\"acker--Williams form $\propto g^2 /k_t^2$, while for 
$\Lambda_{QCD}<<k_t<<\alpha_S\sqrt{\chi}$, the distribution is $\propto 
{1 \over g^2} \log(k_t^2/g^4 N_c\chi(y))$. Even for rapidities far from the 
beam rapidity, it is reasonable to assume that  
$y_{spacetime}\sim y_{mom}$ for wee glue and light sea quarks.  
For $k_T$ very small compared to
$\alpha_S \sqrt{\chi}$, the above formula may break down, but we expect that
a large number of physical quantities will not be infrared sensitive.

While the above expressions provide a very interesting picture of the
behaviour of the gluon distribution as a function of the transverse distance 
$x_t$, the dependence of the distributions on the spacetime rapidity is hidden 
in the function $\chi(\eta,Q^2)$ which has not been specified. In order to 
specify
$\chi$ for all rapidities $\eta$, we need a model of the nuclear color charge
distribution (at large $Q^2$) in the fragmentation region as well. In the next
section, we construct a model for $\chi$ at some reference scale 
$Q_0^2>>\Lambda_{QCD}$. As we will discuss below, $\chi$ obeys evolution 
equations in both 
$y$ and $Q^2$. The important point to note here is that with our model 
of $\chi$ the expression in Eq.~\ref{spaceglue} is then boost covariant. 
This enables us to describe what the wee parton distributions look like in 
any frame.

Finally, the parton distributions computed in the above described manner can 
be related to the empirically extracted gluon distributions at the scale $Q^2$
by the equation
\be
G(x,Q^2)=\int_0^{Q^2} dk_t^2 {dN\over {dx dk_t^2}} \, .
\label{lcglue}
\ee
This simple relation is obtained using light cone quantization and light cone 
gauge $A^+ = 0$~\cite{Basics}.

\section{The rapidity distribution of color charge in nuclei at a fixed 
resolution scale $Q_0^2$}
\vspace*{0.3cm}

In this section we shall formulate a model for $\chi$, the color charge 
squared per unit area per unit rapidity, integrated upto the rapidity of
interest. A similar model for $\chi$, albeit without the explicit $y$ 
dependence, was discussed recently by one of us and 
M. Gyulassy~\cite{gyulassy}. A detailed model in the fragmentation region
is relevant for studying the renormalization group evolution of $\chi$ towards
smaller rapidities and larger $Q^2$. 

Consider a nucleus at rest. For a large nucleus, the color charge squared per 
unit longitudinal distance $z$,
$dC/dzd^2x_\perp$ is a constant within the nucleus when measured on 
longitudinal resolution scales which are large 
compared to the inter--nucleon separation.
The color charge squared is the
typical value of charge fluctuations in the fragmentation region of the 
proton--about 1 unit of rapidity.  Here we must include the fluctuations due to
the gluons as well as the quarks, so this may be roughly a factor of two larger
than is the case for just the valence quarks.
Note also that we get a non zero color charge only because we are choosing to 
measure the charge density on a transverse size scale much less than a fermi.
In a large nucleus, the color charge resolved by the probe corresponds to a 
higher dimensional representation of the color algebra and is therefore 
classical~\cite{RajLar, RajNex}. 
The probe sees only small transverse bits of each nucleon but the
fluctuation fields add incoherently from nucleon to nucleon.

The non--Abelian Weizs\"acker--Williams fields appear as solutions in the 
nuclear rest frame to an external probe in the following sense. 
Outside each nucleon, for the long wavelength small $x$ modes, there
is only the gauge transform of the vacuum field.  As we pass from one
side of a nucleon to the other, we get a different gauge transform
of the vacuum. The field inside a nucleon is not known. Further, since the 
field is pure gauge outside each nucleon,
or in other words, the field strengths are zero, the classical force exerted 
by the long wavelength modes of one nucleon on a neighboring one is negligible.
We therefore have a model of the nucleus with short range nuclear
forces where it is the short wavelength ($x\geq 0.1$, where $x\sim 0.1$ is
the wavelength corresponding to the inter--nucleon separation) modes which are 
responsible for intra-nuclear dynamics such as nuclear binding. This model
is therefore consistent with the conventional picture that nucleonic 
effects in nuclei (such as the EMC effect for instance) are unaffected by
the physics of small $x$ modes in the nucleon 
(for a discussion see for instance Ref.~\cite{Hoyer}). 
The typical change in the Weizs\"acker--Williams field seen by the external 
probe should then be described by a stochastic source with a strength which 
must be determined phenomenologically by an empirical knowledge of the 
nucleon structure functions. It is with this idea in mind that we formulate 
the model below.

In the nuclear rest frame we choose to localize the color charge as
\be
 { {d\tilde{C}} \over {dz}} \propto \theta (z+2R)\theta(-z) \, ,
\label{charge}
\ee
where $\tilde{C}$ is the {\it classical} color charge squared per unit 
transverse area at the scale
$Q_0^2$ of the external probe. In the boosted frame, recall that we define
 $\eta = \eta_{R} + ln(x^-_{R}/x^-)$
where $x^-=(t-z)/\sqrt{2}$ and $x_{R}^-\approx 2R/{\sqrt{2}\gamma}$. 

It is easily seen 
that the above charge distribution is independent of the light cone time
$x^+$ for $x^+<<\gamma$ and $\gamma>>1$. This result is obtained by boosting
the field in the rest frame to the infinite momentum frame. In the classical 
limit we are working in, the field in
one frame is just the Lorentz transform of the field in another. In this
limit, the commutators of the color charges vanish, thereby allowing us to 
relate the charge distributions in the two frames.

The color charge squared per unit transverse area per unit rapidity in the 
nuclear rest frame is 
\be
        {{d\tilde{C}} \over {d\eta}} = { {d\tilde{C}} \over {dz}} {{dz} 
\over {d\eta}} \, ,
\ee
and from our definition in Eq.~\ref{charge}, we have
\be
        {{dz} \over {d\eta}} = 2R  e^{-\eta}\, ,
\ee
for $0<\eta<\infty$.

The color charge squared density in the nuclear rest frame is 
\be
        {{d\tilde{C}} \over {dz}} = \kappa \rho_0
\ee
where $\kappa $ is the typical charge squared {\it per nucleon} in one unit 
of rapidity and $\rho_0$ is the density of nucleons. The latter may be chosen 
for a 
large nucleus to have the typical value for nuclear matter density. All of the
uncertainty about the color charge distribution {\it inside each 
nucleon} is therefore represented in $\kappa$.

We can now boost all this to the frame where the nucleus moves very fast.
The main effect of the boost is to replace the density $\rho_0$ by the nuclear
gamma factor times $\rho_0$.  We find therefore that the
new charge squared per unit area per unit rapidity (which is the
$\mu^2$ introduced in the previous section) becomes
\be
\mu^2 (\eta, Q_0^2) &=& 2 R \kappa \rho_0 e^{\eta_{R}-\eta}\,\,\, ; \, \eta
\geq \eta_R 
\nonumber \\
                 &=& 0 \,\,\, ; \,  \eta<\eta_R \, .
\label{muquation}
\ee
This rapidity distribution is amusing.  In the rapidity variable, most
of the color charge squared is concentrated in one unit of rapidity above
the beam rapidity.  This is a boost invariant statement.  The region
where the charge squared distribution falls to that typical of a single nucleon
is however $\Delta \eta \sim ln(R/1~Fm)$.  Even though most of the
nucleons are concentrated in one unit of rapidity, it takes the order of
$ln(R)$ units in rapidity before the charge squared has fallen to a small 
value!

It is also interesting to consider how this picture of the fragmentation
region can be formulated in terms of the momentum space rapidity. A similar
argument to the one above applies if we formulate our discussion in
terms of a momentum space rapidity that we define as  $y=
y_{R} + \log(k^+/P_{R}^+)$. Note that this definition is 
not the usual one since $P_{R}^+$ is {\it not} the nuclear momentum per
nucleon $P^+$, but instead the momentum that is conjugate to the Lorentz
contracted width $x_{R}^-$ of the nucleus. This definition has the
following advantages. Firstly, from the uncertainty principle, it is 
equivalent to the space--time rapidity we defined previously since 
$\log(k^+/P_{R}^+)=\log(x_{R}^-/x^-)$. Secondly, by this definition,
momenta $P_{R}^+<k^+<P^+$ correspond to fragmentation region rapidities, or
in terms of $x^-$, wavelengths shorter than the Lorentz contracted width
of the nucleus. One therefore expects that the charge distribution in terms
of the above defined momentum space rapidity is identical to the distribution in Eq.~\ref
{muquation}: \, $\chi(\eta,Q^2)\equiv \chi(y,Q^2)$. 
What fraction $x_{R}$ of the nuclear momentum $P^+$ does
$P_{R}^+$ correspond to?
It may be determined by requiring that 
it correspond to a wavelength equal to the Lorentz contracted width of
the nucleus $1/P_{R}^+ = x_{R}^-$. Hence,  
\be
{1\over {x_{R}P^+}}\sim {\sqrt{2} R\over {P^+/m_N}}\Longrightarrow 
x_{R}\sim A^{-1/3}/(\sqrt{2}\, r_0 m_N) \, ,
\label{waves}
\ee
where $m_N$ is the nucleon mass and $r_0=1.12$ fm. For 
$A=200$, this corresponds to $x_{R}\sim 0.025$. Therefore, 
by our definition, for $A=200$, fragmentation region rapidites 
${\tilde y}>{\tilde y}_{R}$ correspond to 
light cone momentum fractions $x>0.025$. Of course even for these large
nuclei our classical considerations are very marginal and can be safely
applied in the theoretical limit of much larger nuclei or equivalently,
when the parton densities in the fragmentation region are already large.

Returning to Eq.~\ref{muquation}, because most of the color charge squared is 
contained in one unit of rapidity,
there is not much dynamical evolution of the glue distribution
due to the glue from different nuclei interacting in the fragmentation
region.  This is because the typical interaction strength is on the order of
$\alpha_S$ which is small due to the high field density and because the
typical distance it takes to evolve the gluon distribution due to the 
interactions of the gluons among themselves is of order $1/\alpha_S$.
Treating the nucleons as a source of glue 
in the nuclear fragmentation region is therefore justified so long as 
sources due to gluons of order one unit of rapidity from the beam rapidity 
are added to those from the valence quarks.  This of course introduces a 
phenomenological parameter ($\kappa$ above) which we cannot precisely 
determine from first principles. However, as will be discussed below $\kappa$ 
can be determined empirically from the nucleon gluon distributions.

The gluon rapidity distribution varies slowly ($\propto \alpha_S$ per unit 
rapidity)  when $\Delta y << 1/\alpha_S$
away from the beam rapidity.  It is obtained by solving the 
classical equations of motion for the source distribution above.
Integrating the expression in Eq.~\ref{muquation} above, we obtain
\be
        \chi (y, Q_0^2) &=& 2\, R \rho_0 \kappa \, ; \, y<y_R \nonumber 
\\ 
&=&  2 \,R\rho_0 \kappa e^{y_R-y}\, ; \, y>y_R
\, , 
\label{chigrv}
\ee
where, we remind the reader, $y_R$ is the momentum space rapidity defined 
previously. The ratio of 
the nuclear gluon distribution to the nucleon parton distribution times 
A in the nuclear fragmentation region for very large nuclei is plotted in
Figure 1. Our classical considerations result therefore in a charge
distribution that is independent of rapidity for $y<y_R$. 
Since any 
gauge invariant configuration of fields must satisfy Gauss's law, the 
contribution of glue at small x is necessarily associated with the valence 
degrees of freedom and therefore also exhibits boost invariance. This is also
clear from examining Eq.~\ref{fourier}.

\section{Discussion}
\vspace*{0.3cm}

From our elementary considerations above, we have obtained an 
analytic, boost covariant form for the color charge squared distribution 
$\chi$ in the 
nuclear fragmentation region at the momentum resolutution scale $Q_0^2$. 
It relies on one parameter $\kappa$ which encapsulates our uncertainty about 
the non--perturbative physics in the fragmentation region. However, since 
$\kappa$ has the
transparent physical interpretation of being the color charge squared per 
nucleon, some of the uncertainty can be eliminated by expressing $\kappa$ in 
terms of the nucleon gluon density. Hence, at some reference scale $Q_0^2$, 
\be
\kappa  = {1\over {2\,N_c}} \int_{\Delta y\sim 1} dx \left(
V(x,Q_0^2)+S(x,Q_0^2)\right)
+ {N_c\over {N_c^2-1}} \int_{\Delta y\sim 1}dx G(x,Q_0^2) \, .
\label{defkap}
\ee
Here $V$, $S$ and $G$ denote respectively the valence, sea quark and
gluon densities. They are integrated over the roughly one unit of rapidity
that one estimates the nucleons to occupy in the fragmentation region.
The factors multiplying the parentheses above are the color charge squared 
per quark and the color charge squared per gluon respectively. The last term 
for example, has the interpretation that it is the color charge squared per 
gluon times the number of gluons in the nucleon contained in one unit of 
rapidity in the fragmentation region. Similarly so for the first term.

The parton densities at the reference scale $Q_0^2$ can be parametrized by 
fits to the nucleon deep inelastic scattering data. A convenient one is the 
NLO parametrization of Gl\"uck, Reya and Vogt~\cite{GRV}.
In principle, the sensitivity of the results to different parametrizations 
can be investigated. As is well known by now, the GRV parametrizations of the 
data at the (extremely low!) reference scale ${(Q_0^{grv})}^2=0.34$ GeV$^2$, 
evolved 
with the next to leading order Altarelli--Parisi evolution equations, agree 
well with the DIS data from HERA. The GRV parametrizations
have the generic form 
\be
xf(x,Q^2)=Nx^a\left(1+Ax^b+Bx+Cx^{3/2}\right)(1-x)^D \, ,
\ee
where $f=V,S,G$ and the coefficients and powers in the above expression are 
polynomials of 
$\log(\log(Q^2/\Lambda_{QCD}^2))$. The detailed expressions are given in 
Ref.~\cite{GRV}. 
For our classical considerations to apply, we require the reference scale 
$Q_0>>{Q_0^{grv}}\sim \Lambda_{QCD}$. 
The parton densities which determine $\kappa$ can then be parametrized by
the GRV parton distribution evolved to the scale of interest $Q_0$.

With the initial data for $\chi$ as input, the Wilson renormalization 
arguments of Ref.~\cite{AlexLar} can be used
to determine it at higher $Q^2$ and smaller values of $y$. 
It has the form
\be
{d\chi\over {dy dQ^2}} = {N_c\over{N_c^2-1}} {1\over {\pi R^2}} 
{dN\over {dy dQ^2}} \, ,
\label{Wilson}
\ee
where $dN/dydQ^2$ is the distribution function in Eq.~\ref{fourier}.
This renormalization group equation can be formulated in DGLAP form by first 
integrating over y,
\be 
{d\chi\over {dQ^2}} = {N_c\over {N_c^2 -1}} {1\over {\pi R^2}}
\int_y^{\infty} dy^\prime {dN\over {dy^\prime dQ^2}} \, .
\label{Wilsonp}
\ee
Integrating over $Q^2$, we obtain the BFKL--like form
\be
{d\chi\over {dy}}={N_c\over {N_c^2 -1}} {1\over {\pi R^2}}
\int_0^{Q^2} d{Q^\prime}^2 {dN\over {dy d{Q^\prime}^2}} \, .
\ee
It was shown recently that for large momenta the kernel of
the equation satisfied by the right hand side of the above is precisely the 
BFKL kernel~\cite{Kovner}. However 
in general, as discussed earlier, the gluon distribution on the right hand 
side of the above equations has a non--linear dependence on $\chi$.

A clear picture of shadowing then emerges in the above model. From the
perturbative QCD standpoint, one of the possible sources of shadowing is
 from the the input parton distributions 
which are modified upon $Q^2$ evolution by the splitting functions a la Altarelli--Parisi~\cite{Kumano,Eskola}.
The other source is due to recombination and screening effects which enter 
into the evolution equations as higher twist effects~\cite{GLR,MuellerQiu}. 
There 
have been a number treatments of shadowing with take both these effects into 
account~\cite{Qiu,Eskola,EsQuWa}. It is often believed that current data
can be accounted for by shadowing of the initial (non--perturbative) 
parton distributions and leading twist DGLAP evolution of these
distributions.

On the other hand, our model in the previous section suggests that for 
sufficiently large nuclei i) the initial parton distributions can be 
determined perturbatively (modulo $\kappa$) and ii) they are not shadowed. 
However the shadowing away from the fragmentation region is large due to 
higher twist effects which, in this model, are already present at the 
classical level. If we substitute the classical result of Eq.~\ref{fourier} 
into the right hand side of Eq.~\ref{Wilsonp} 
\be 
{d\chi\over {dQ^2}} = {1\over {\pi^2 \alpha_S}}
\int_x^{1} {dx^\prime\over x^\prime} \int_0^{1\over \Lambda_{QCD}} {dz
\over z} J_0 (Q\,z)\left[ 1-\exp\left(\beta\log(z\Lambda_{QCD}\right)\right]
\, ,
\ee
where $\beta = 2\pi \alpha_S^2 N_c \chi(x^\prime,Q^2)z^2$. If we expand out
the $1-exp(\cdots)$ term on the right hand side and make the identification
$\chi={N_c\over {N_c^2-1}}x G(x,Q^2)/\pi\,R^2$, we obtain an
evolution equation in the gluon density $G(x,Q^2)$ which on the right hand
side includes all powers of $G(x,Q^2)/{Q^2 R^2}$--or all twists. The first
term in the expansion is just the usual DGLAP evolution equation at small x.
A very similar expression to the above has been obtained by Levin and 
collaborators working in the nuclear rest frame--and is the perturbative
QCD form of the Glauber--Gribov--Mueller formula of high energy 
scattering~\cite{Ayala}. Since the second term in the expansion above can
be related to diffractive scattering, diffractive scattering data can
be used to determine the degree of nuclear shadowing in this approach. 
This was done recently by Capella et al. who used HERA diffractive data
on e--p scattering to reproduce shadowing data from fixed target
experiments--thereby providing a nice check of the Glauber--Gribov--Mueller
expression upto second order in the gluon density~\cite{Capella}.

Ultimately, when parton densities are large, the quantum corrections to
the above formula will be large and have to be computed. For instance, the
second term in the above equation gives only the recombination 
(or diffractive) 
contribution to shadowing (with positive sign!). 
The screening corrections will appear when
one includes quantum corrections. If the well known AGK sum rules~\cite{AGK}
apply, then as argued by Mueller and Qiu~\cite{MuellerQiu}--
the expression will be the same as the above but with a negative sign! This 
has to be confirmed in our approach before anything quantitatively new can be 
added to current 
phenomenology. Moreover,  we hope that in this effective action approach it 
will be easier to compute higher order terms as well~\cite{JKW}.

\section{Summary and Outlook}
\vspace*{0.3cm}

In this paper we have formulated a simple model for $\chi(y,Q_0^2)$, the
color charge squared integrated from the rapidity of interest upto the
beam rapidity at some reference scale $Q_0^2$. At this scale, $\chi$ is the
color charge in the nuclear fragmentation region. $\chi$ depends on the non--
perturbative physics at this scale through $\kappa$ which is the charge squared
per nucleon and is simply related to the {\it nucleon} parton distributions. 
The expression derived for $\chi$ is boost covariant. The Wilson 
renormalization group equation 
discussed in Ref.~\cite{AlexLar} can be used to evolve $\chi$ to $Q^2>Q_0^2$ 
and/or rapidities away from the fragmentation region. We have argued that
this classical picture provides important insight into the nature of 
perturbative gluon shadowing and have argued in the previous section that our
results are closely related to similar approaches in the nuclear rest frame. 
This is indeed as it should be because the model we have constructed is
boost covariant. The boost covariant gluon
distributions thereby constructed can be applied to a wide range of 
phenomenology including gluon and quark
shadowing~\cite{FranStrik,Arneodo}, intrinsic quark production~\cite{RajLar3} 
and nuclear collisions at ultrarelativistic energies~\cite{KLW,gyulassy}. 
It is however important to understand the quantum corrections in the
infinite momentum frame effective action picture before significant additions
can be made to current phenomenology~\cite{JKW}.

\section*{Acknowledgments}

This work was initiated at the INT, Seattle during its fall 1996 program
``From structure functions to the quark gluon plasma in ultrarelativistic 
nuclear collisions''. We would both like to thank the INT faculty and staff
for their support and all the program participants for stimulating 
discussions.
In particular, we would like to thank Rajiv Gavai, Miklos Gyulassy, 
Jamal Jalilian--Marian, Yuri Kovchegov,  
Alex Kovner, Andrei Leonidov, Berndt M\"uller, Dirk Rischke
and Heribert Weigert for discussions relevant to this paper. Our research is 
supported by the U.S. Department of Energy under grants
No. DOE Nuclear DE--FG06--90ER--40561 and  DE-FG02-87ER40328 and the 
Danish Research Council.

\section{Figure caption}
\vspace*{0.3cm}

{\bf Fig. 1}: Ratio $R_A^{frag}=x G_A(x,Q_0^2)/A \kappa$ (where $\kappa$ is 
defined as in Eq.~\ref{defkap}) as a function of rapidity in the
fragmentation region. Above $y_R$ is defined by Eq.~\ref{waves}; $y_{beam}-
y_R \sim \log(A^{1/3})$ and $Q_0^2>>\Lambda_{QCD}^2$.

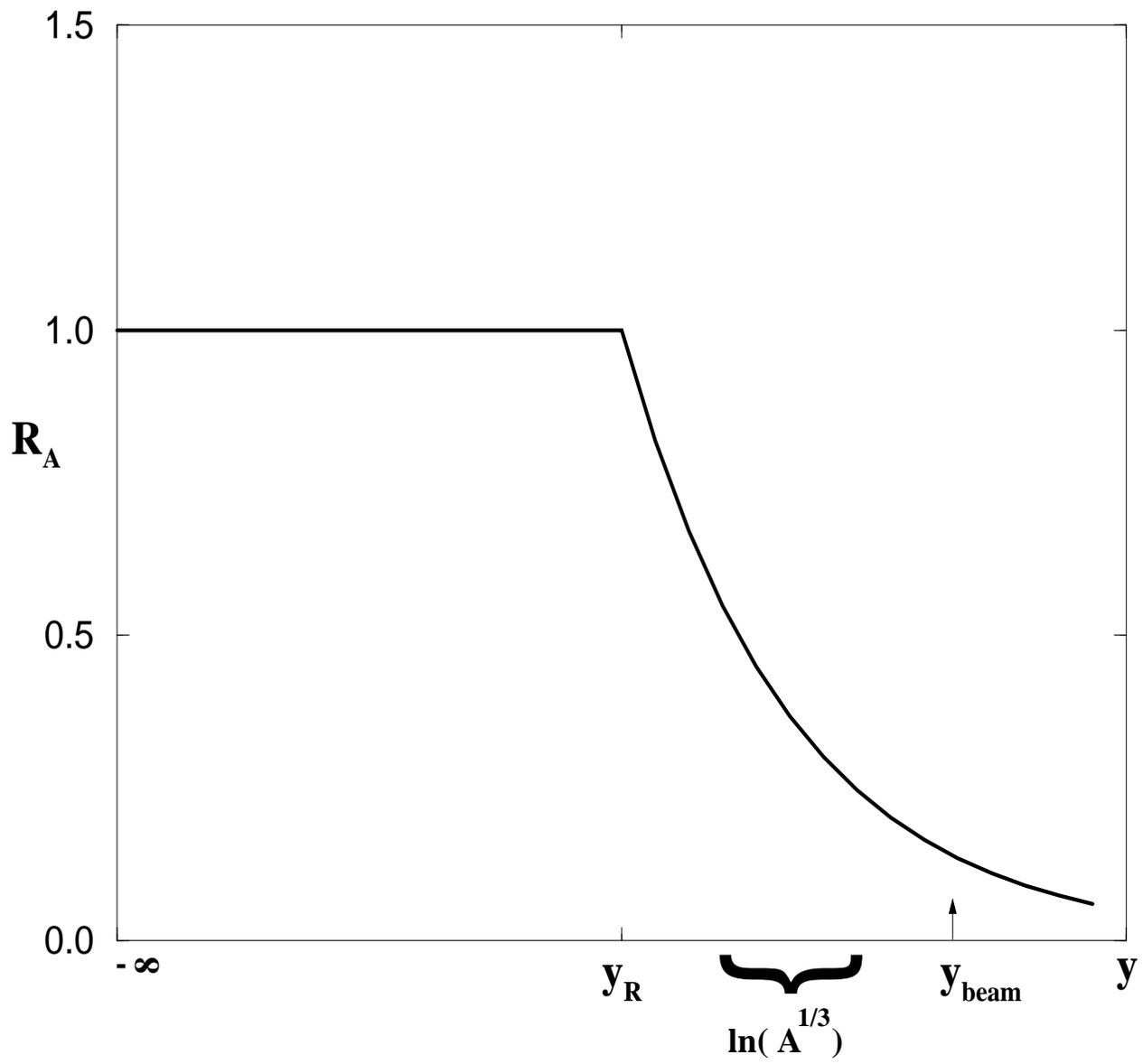
\end{document}